\documentclass[sigconf]{acmart}

\usepackage{subcaption}
\usepackage{graphicx}
\usepackage{fancyhdr}

\AtBeginDocument{%
  \providecommand\BibTeX{{%
    \normalfont B\kern-0.5em{\scshape i\kern-0.25em b}\kern-0.8em\TeX}}}

\setcopyright{none}
\renewcommand\footnotetextcopyrightpermission[1]{} 
\makeatletter
\renewcommand\@formatdoi[1]{\ignorespaces}
\makeatother

\acmBooktitle{Generative AI and HCI Workshop – CHI ’22, May 10, 2022, New Orleans, LA} 

\begin{document}

\title{Team Learning as a Lens for Designing Human--AI Co-Creative Systems}

\author{Frederic Gmeiner}
\affiliation{%
    \institution{HCI Institute}
    \institution{Carnegie Mellon University}
    \city{Pittsburgh}
    \state{PA}
    \country{USA}}
\email{gmeiner@cmu.edu}

\author{Kenneth Holstein}
\affiliation{%
    \institution{HCI Institute}
    \institution{Carnegie Mellon University}
    \city{Pittsburgh}
    \state{PA}
    \country{USA}}
\email{kjholste@cs.cmu.edu}

\author{Nikolas Martelaro}
\affiliation{%
    \institution{HCI Institute}
    \institution{Carnegie Mellon University}
    \city{Pittsburgh}
    \state{PA}
    \country{USA}}
\email{nikmart@cmu.edu}

\renewcommand{\shortauthors}{Gmeiner, et al.}

\begin{abstract}
  Generative, ML-driven interactive systems have the potential to change how people interact with computers in creative processes – turning tools into co-creators. However, it is still unclear how we might achieve effective human--AI collaboration in open-ended task domains. There are several known challenges around communication in the interaction with ML-driven systems. An overlooked aspect in the design of co-creative systems is how users can be better supported in \textit{learning to collaborate} with such systems. Here we reframe human--AI collaboration as a learning problem: Inspired by research on \textit{team learning}, we hypothesize that similar learning strategies that apply to human-human teams might also increase the collaboration effectiveness and quality of humans working with co-creative generative systems. In this position paper, we aim to promote team learning as a lens for designing more effective co-creative human--AI collaboration and emphasize \textit{collaboration process quality} as a goal for co-creative systems. Furthermore, we outline a preliminary \textit{schematic framework} for embedding \textit{team learning support} in co-creative AI systems. We conclude by proposing a research agenda and posing open questions for further study on supporting people in learning to collaborate with generative AI systems. 
\end{abstract}



\settopmatter{printfolios=true}

\maketitle

\section{Introduction}

Generative ML-driven systems introduce novel ways to interact and create with digital tools. 
With the ability to automate tedious tasks or generate novel and surprising content, such systems are increasingly shifting from a role as passive tools towards becoming active and autonomous agents working with human creators --- making the tool a collaborator in a co-creative process \cite{guzdial_interaction_2019}. 
 
Generally, effective human-human collaboration is grounded in communication and mutual understanding \cite{resnick_grounding_1991}. However, with various open questions around effective communication, ML-driven generative tools pose many challenges to human--AI collaboration. 

The field of \textit{team learning} (or group learning) emerged from within organizational sciences by studying how human groups learn to collaborate effectively together. Despite the lack of a uniform definition of the term team learning, two key concepts are commonly attributed to establishing effective teamwork: A \textit{grounding in communication} and building \textit{shared mental models (SMMs)} among team members \cite{resnick_grounding_1991,mohammed_team_2001}. Developing SMMs involves adaptive coordination of actions among team members, along with developing shared representations of the task and learning each other’s abilities, limitations, goals, and strategies \cite{van_den_bossche_team_2011, dechurch_measuring_2010, fiore_technology_2016, scheutz_framework_2017}.

Recent work has argued for drawing inspiration from such known factors that make human-human collaboration effective for designing more effective human–AI collaborative systems \cite{zhou_group_2018, cai_hello_2019, bittencourt_conceptual_2020}. For example, \textit{Kaur et al.} \cite{kaur_building_2019}, argue for supporting the development of shared mental models as a foundation for \textit{effective collaboration} among human–-AI teams. 

Even though collaboration plays an integral aspect in how users interact with co-creative AI systems, it is often considered a secondary quality in the system's design. Most systems are evaluated based on the quality of their output, such as an interesting melody or an optimized component that meets engineering requirements while minimizing weight. However, it is less common to evaluate systems based on \textit{process-oriented} measures of collaboration quality; for example, \textit{Guzdial et al.} studied specifically whether collaboration with different co-creative AI systems was experienced as enjoyable or frustrating for game designers. Moreover, \textit{Kreminski and Mateas} promote the idea that the design of co-creative systems should \textit{"prioritize the aesthetic experience of the creative process over that of the resulting product"} \cite{kreminski_reflective_2021}. Other authors such as \textit{Bown and Brown} \cite{filimowicz_interaction_2018} or \textit{Kantosalo et al.} \cite{kantosalo_modalities_2020} emphasize human-centered interaction design to enhance the collaboration \textit{experience} of co-creative systems.    

\begin{figure*}
    \includegraphics[width=\textwidth]{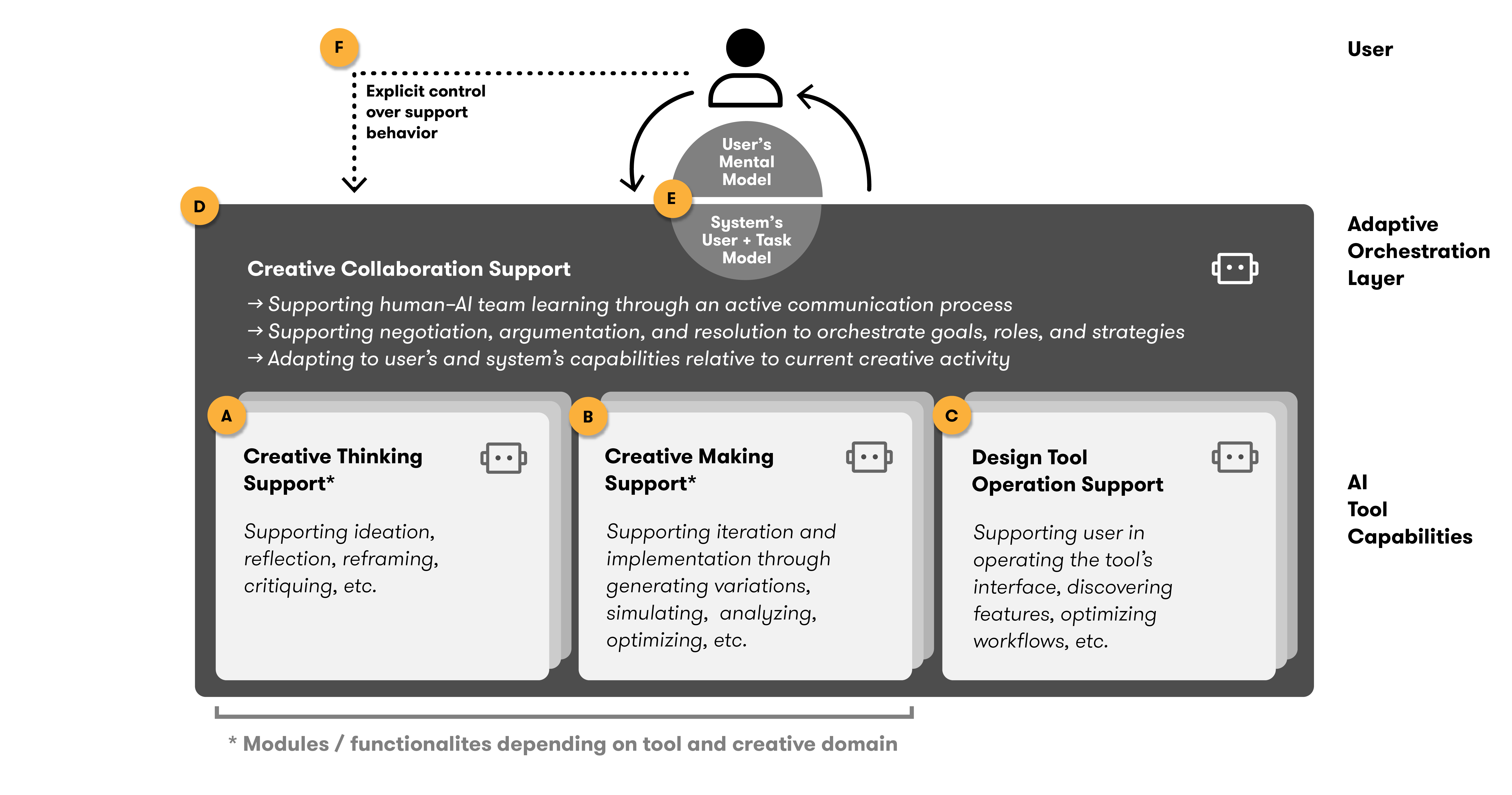}
    \caption{Diagram illustrating the schematic framework for supporting co-creative human--AI team learning comprising (A-C) core capabilities supporting creative processes and tool usage, and (D) adaptive support layer for orchestrating creative collaboration between user and system by building (E) shared mental models. (F) Users have explicit control over the system's support behavior.}
    \label{fig:support-layers}
\end{figure*}

We follow these endeavors and propose positioning the \textit{quality of the creative collaboration process} as an important driving design principle of co-creative systems. Furthermore, to facilitate effective collaboration between user and system, we argue that a key design strategy lies in actively supporting users in learning how to collaborate with co-creative systems --- reframing human--AI collaboration as a \textit{team learning} problem.
Our motivation is driven by the way heterogeneous human teams learn to effectively collaborate through an active process of negotiation, "constructive" conflict, argumentation, and resolution \cite{head_effective_2003,jeong_knowledge_2007,van_den_bossche_team_2011}. 
Hence, instead of users learning to interact with a co-creative system as a separate, explicit task, we envision this learning support to become an \textit{essential} part of a co-creative system's \textit{behavior}. 

Systems can assist users in learning how to work co-creatively by helping users hone their mental models of the relative capabilities of the system versus their own. This involves creating a self-awareness on the user's side about their own needs, capabilities, and limitations while also helping users understand the capabilities and limitations of the co-creative system.
At the same time, to effectively support the user, the system needs to hone its mental model of the user's goals, capabilities, and limitations regarding the current creative task at hand --- by maintaining an active communication process, explicitly and implicitly.  

With this position paper, we contribute the following:
(1) We promote team learning strategies as a lens for supporting effective human--AI co-creation.
(2) We share the outlines of a schematic framework for supporting users in learning to co-create with generative AI systems as a springboard for further exploration, discussion, and refinement through the research community.
We conclude by sketching out a research agenda and formulating open questions to stimulate future discussions around the design of team learning support systems within co-creative human--AI applications.

\section{Towards team learning support for human--AI co-creation}

In Figure 1, we outline a schematic framework illustrating how co-creative AI tools could be designed to support users in effectively collaborating with the system based on team learning principles. 

In this framework, modules \textit{A -- C} represent possible core functionalities of the system. The functionalities of the category \textbf{(A)} support conceptual tasks such as ideation and critical thinking about a creative activity, which we summarize as \textbf{creative thinking support}. Examples of creative thinking support include \textit{ImageSense} \cite{koch_imagesense_2020}, a system by \textit{Koch et al.} that supports designers' mood boarding process by suggesting topic-related semantic concepts, or systems capable of stimulating reflection about a design process, like \textit{ReflectionSpace} \cite{sharmin_reflectionspace_2013} by \textit{Sharmin and Bailey}. 

Furthermore, systems can also directly support \textbf{(B)} the production of creative outputs by generating artifacts such as images, text, music, or 3D geometries – summarized here as \textbf{creative making support}. Such systems' functions contribute to design iteration and implementation processes by generating outputs based on user-defined goals and constraints; for example, supporting the creation of 3D structures out of actively transforming materials like \textit{SimuLearn} by \textit{Yang et al.} \cite{yang_simulearn_2020}. 

In addition to supporting dedicated domain tasks, another core functionality of a system concerns supporting general tool operation and learning about task-relevant features in general \textbf{(C)}. This \textbf{design tool operation support} includes helping users operate the tool interface, discover relevant features of the task, or improve user workflows, for example, using adaptive interface techniques such as \textit{command disambiguation} \cite{lafreniere_these_2015} or \textit{interactive guiding systems} \cite{fernquist_sketch-sketch_2011}.

However, as of today, most existing generative systems are designed to support a narrow range of creative tasks \cite{frich_mapping_2019}. Nevertheless, we imagine that systems will increasingly unite a multitude of different and complementary generative capabilities in a modular fashion to augment a broader range of creative processes and phases. Hence, while \textbf{A} and \textbf{B} do not cover all phases of a creative process, they are presented here as simplified categories in order to \textit{highlight key opportunities} for designers of co-creative systems to identify additional creative support functionalities complementing the already present capabilities of a given generative system. However, in reality, these will likely vary depending on the domain and the overall functionality of the co-creative tool.

Atop these functional capabilities, we imagine features based on team learning principles that provide \textit{adaptive support for orchestrating the collaboration} between the user and the tool's capabilities. This \textbf{(D)} \textbf{creative collaboration support} acts like a team coach --- establishing an active process of communication between user and system, allowing for negotiation, argumentation, and resolution of goals, roles, and strategies while considering the user's and system's capabilities and limitations in relation to the creative task.

Since every user has individual needs, capabilities, and preferences (which also constantly change according to project and task), the behavior of the collaboration support agent will likely be highly adaptive. For example, an industrial designer might be experienced in using a generative design tool for shape exploration, with the co-creative tool suggesting shape variations and alternative design strategies or goals. Here, the designer and the system know how to co-create together and require only minor team learning support. But at the same time, the designer might be less knowledgeable about manufacturing processes and unaware of the tool's manufacturing support features. In this situation, the system could raise the user's awareness of the existing support feature to effectively complement each other's capabilities.    

To build and maintain \textbf{shared mental models (E)}, the system would derive its user model through explicit and implicit mechanisms; for example, asking the user for information about the specific design task (such as parsing a written design brief to understand the context and goals of the user \cite{tan_text2scene_2019}), or observing the user’s behavior to implicitly infer the user’s design goals \cite{law_design_2020}. 
Furthermore, the collaboration support should not be designed as an autonomous black box, but rather be \textbf{transparent to the user and allow direct control} over it \textbf{(F)}. For example, if users do not prefer to work collaboratively with the tool, they should be able to minimize the provided team learning support.  

There are still many unknowns for building such an adaptive support system. 
Consequently, this conceptual framework serves as a starting point for further exploration, discussion, and refinement. Nevertheless, we think there is a growing need and potential to support human--AI co-creation through team learning strategies to increase collaboration quality.
The following section outlines our planned research agenda to explore how human--AI team learning can be facilitated in co-creative task domains. 

\section{Research Approach}

Based on the formulated framework, we identified relevant areas and a research agenda to better understand how human--AI teams can learn to collaborate in co-creative task domains.

As a first step, our objective is to gain a better understanding of how professional users learn to collaborate with existing generative AI tools (including both research prototypes and commercial tools) and what obstacles users face \textit{without} dedicated collaboration learning support. Therefore, we plan to conduct formative think-aloud studies \cite{van_someren_think_1995} observing designers from different domains working with generative systems on real-world tasks. This will allow us to identify existing support functionalities and learning strategies of co-creative systems in relation to additional currently missing complementary support functionalities (\textit{Figure 1} \textit{A–C}) of generative systems and user types.

Furthermore, in some of our studies, we will simulate an AI support system by having a human expert supporting the user. From these guided sessions, our aim is to gather insights for the design of a team learning support system (\textit{Figure 1}\textit{D}).
For example, observing the support strategies, pedagogical moves, and communication patterns of the human expert guide can provide insight into when and how to guide users in learning to better co-create with the AI system. Consequently, we would analyze the video and screen recordings of the single and guided think-aloud sessions by applying measures inspired by strategies for assessing human-human collaboration \cite{burkhardt_approach_2009, meier_rating_2007} and team learning \cite{rummel_learning_2009}. 

Through these activities, we intend to identify situations and opportunities related to effective communication and the development of shared mental models (\textit{Figure 1}\textit{E}) -- such as modeling the user's state and goals through explicit or implicit methods or fostering the user’s self-awareness of their own capabilities in relation to the capabilities of the system. Furthermore, we will collect data on the general attitudes of the user towards AI-driven co-creation and collaboration preferences relevant to agency, control, and team roles (\textit{Figure 1}\textit{F}).

To direct our observations on learning to co-create with generative systems in professional contexts (and not on learning software tools in general), we would recruit professional participants who are proficient in working with their domain-specific software applications (Computer-Aided Design, Digital Audio Workstation, Video Editing Suite, etc.) but ideally have no prior experience working with generative AI tools in their workflow. 

Following the formative observations, we would apply the derived insights to design novel interfaces and interactions for supporting human--AI co-creation. This would involve iterative prototyping of intelligent user interfaces using Wizard-of-Oz techniques to quickly test different collaboration and support strategies through explicit or implicit communication.

To draw conclusions on the collaboration performance, we would evaluate the effectiveness of the developed prototypes in supporting co-creative human--AI collaborations by similar user studies as outlined above, allowing a comparison between all three conditions (without support, with human support, and with AI support).

On a broader level, we envision that these activities will help identify \textit{learning goals} and \textit{develop guidelines} to inform the design of future interfaces for generative AI tools that facilitate human--AI collaboration. Specifically, we would like to raise the following questions for further discussion: 
\begin{itemize}
    \item Is team learning a desirable approach for all users and creative domains?
    \item What are possible team learning goals of co-creative systems?
    \item What modalities and communication styles support nondisruptive in-action team learning during flow-state co-creative activities? 
\end{itemize}

With the work presented in this paper, we hope to inspire other researchers to add to the discussion of team learning in co-creative AI systems and to contribute to the ongoing effort of identifying and developing models for realizing effective collaboration between humans and computer systems in creative task domains.

\section{Conclusion}
Co-creative, ML-driven generative systems challenge effective communication and collaboration between users and computer systems. To address this issue, we introduce \textit{team learning} as a lens for designing effective human--AI co-creative systems by taking inspiration from how human-human teams learn to collaborate through active communication processes such as negotiation, argumentation, and resolution to form \textit{shared mental models}. Concretely, we present a \textit{schematic framework} (\textit{Figure 1}) that utilizes team learning strategies to support users in learning to co-create with generative AI systems. Finally, we propose a \textit{research agenda} to further explore co-creative human--AI team learning support. We anticipate that this work will improve co-creative software, enhance people's creative capabilities, and generally lead to more effective human--AI collaboration in open-ended task domains. 

\section{Acknowledgments}

This material is based on work supported by the National Science Foundation under Grant No. 2118924. Any opinions, findings, and conclusions or recommendations expressed in this material are those of the author(s) and do not necessarily reflect the views of the National Science Foundation.

\bibliographystyle{ACM-Reference-Format}


\end{document}